\documentclass[prd,nofootinbib,onecolumn,english,cp1251,T2A,showpacs,showkeys]{revtex4}
\usepackage{babel}

\usepackage{amssymb}
\usepackage{amsmath}
\usepackage{slashed}

\usepackage[dvips]{graphicx} 

\usepackage{amsthm} 

\usepackage{varioref}   
\usepackage{xr}         

\usepackage{pstricks}           

\allowdisplaybreaks[1]

\newcommand\nn{\nonumber}
\newcommand\ba{\begin{eqnarray}}
\newcommand\ea{\end{eqnarray}}
\newcommand\baf{\begin{eqnarray}}   
\newcommand\eaf{\end{eqnarray}}
\newcommand\eq[1] {\begin{align} #1 \end{align}}   

\newcommand{\br}[1]{\left( #1 \right)}
\newcommand{\brs}[1]{\left[ #1 \right]}
\newcommand{\brf}[1]{\left\{ #1 \right\}}

\newcommand{\vv}[1]{{\bf #1}}

\newcommand{\GeV}{\mbox{GeV}}

\begin{document}

\title{A new conception in describing the Drell-Yan type processes}

\author{A.~I.~Ahmadov}
\email{ahmadov@theor.jinr.ru}
\affiliation{Joint Institute for Nuclear Research, 141980 Dubna, Moscow Region, Russia}
\affiliation{Institute of Physics, Azerbaijan National Academy of Science, Baku, Azerbaijan}

\author{Yu.~M.~Bystritskiy}
\email{bystr@theor.jinr.ru}
\affiliation{Joint Institute for Nuclear Research, 141980 Dubna, Moscow Region, Russia}

\author{E.~S.~Kokoulina}
\email{kokoulin@sunse.jinr.ru}
\affiliation{Joint Institute for Nuclear Research, 141980 Dubna, Moscow Region, Russia}

\author{E.~A.~Kuraev}
\email{kuraev@theor.jinr.ru}
\affiliation{Joint Institute for Nuclear Research, 141980 Dubna, Moscow Region, Russia}

\date{\today}

\begin{abstract}
    A set of evolution equations for correlators of densities of quark and gluons
    is considered. Approximate solutions is obtained in frames of gluon and quark dominance.
    A new formulation of the cross sections of Drell-Yan process is suggested.
    Differential cross sections for the QCD sub-processes of type $2 \to 2$ are obtained.
    Sub-process $g b\to tH^-$ as well considered.
\end{abstract}



\maketitle

\section{General Formalism}

The quark parton model of Feynman \cite{Feynman:1973xc} provides the simple description of deep inelastic
phenomena as well as Drell-Yan processes. Theoretical justification of this model was done in terms of
asymptotically free gauge theories \cite{Politzer:1974fr}. Drell-Yan picture based on factorization of
contributions from small and large distances was justified in papers of Collins \cite{Collins:1989gx}.
Deviation from the naive Bjorken scaling of the structure functions of deep inelastic scattering (DIS) was recognized to be
broken by the so called "large logarithms" --- the logarithms of the ratio of momentum squared $Q^2=-q^2$
(virtualities) of particles deep off-mass-shell to their masses. The reasons of appearing of these
logarithms in QED was clarified by methods of quasi-real photons and electrons \cite{vonWeizsacker:1934sx, Williams:1934ad, Kessler:1960, Baier:1973ms}.
Keeping in mind the contributions of higher orders of perturbation theory,in the leading logarithmical
approximation description of processes with large virtualities can as well be formulated in parton
language with some definite dependence on $Q^2$ of partons (quark, gluons) densities --- structure functions,
$q^i\br{x,t}$, $G\br{x,t}$.

The evolution equations of Altarelli--Parisi (AP) \cite{Altarelli:1977zs} for these densities have a form
\ba
    \frac{dq^i\br{x,t}}{dt} &=&
    \frac{\alpha_s\br{t}}{2\pi}
    \int\limits_x^1 \frac{dy}{y}
    \brs{
        q^i\br{y,t} P_{qq}\br{\frac{x}{y}} + G\br{y,t} P_{i G}\br{\frac{x}{y}}
    },
    \qquad
    t = \ln\br{\frac{Q^2}{m^2}},
    \label{DGLAP}
    \\
    \frac{dG\br{x,t}}{dt} &=&
    \frac{\alpha_s\br{t}}{2\pi}
    \int\limits_x^1 \frac{dy}{y}
    \brs{
        \sum_{i=1}^{2N_f} q^i\br{y,t} P_{G q^i}\br{\frac{x}{y}} + G\br{y,t} P_{GG}\br{\frac{x}{y}}
    },\quad i=u,d,s,c,b,t. \nn
\ea
They describe the dependence of quarks ($q\br{x,t}$) and gluons ($G\br{x,t}$) densities with the energy fraction $x$ inside the proton
on a scale of $Q^2$ (where $Q^2>0$). Here $m$ is the suitable normalization point $m \sim Q_0 \sim M_p \approx 1~\GeV$,
$\alpha_s\br{t}$ is the QCD coupling constant on the scale $Q^2$ and
$P_{ij}\br{x}$ are the AP equation kernels:
\ba
    P_{qq}\br{z} &=&
    C_F \br{\frac{1+z^2}{(1-z)}_+ + \frac{3}{2}\delta\br{1-z}}, \qquad
    \nn
    \\
    P_{Gq} &=& C_F\frac{1+(1-z)^2}{z}; \qquad P_{qG}=\frac{1}{2} \brs{z^2+(1-z)^2}; \nn \\
    P_{GG}\br{z} &=&
    2 C_V \br{ \frac{1-z}{z} + \frac{z}{(1-z)_+} + z\br{1-z} + \frac{11}{12} \delta\br{z-1}},\qquad
    \nn
\ea
with $C_F=\frac{N^2-1}{2N}$ and $C_V=2N$ for the color group $SU(N)$.
These quantities satisfy the following properties:
\ba
    \int\limits_0^1 dz~P_{qq}\br{z} = 0,
    \qquad
    \int\limits_0^1 dz ~z~P_{GG}\br{z} = 0.
    \label{PProperties}
\ea

It's useful to remind here the statistical interpretation of AP equations in terms of densities \cite{Lipatov:1974qm}
by means of a set of correlation functions, satisfying the system of statistical equations (renormalization group equation).
It was a success of the numerous applications of APL set of equations, working with two densities $q, G$ of
quarks and gluons into a proton.

Problems associated with processes with large multiplicity \cite{Kokoulina:2004gu, Kokoulina:2005xx, Ermolov:2005tk}, however require some generalization of traditional approach,
introducing the correlation functions \cite{Bukhvostov:1985rn}.

Namely let us introduce $D^q(x,t)$, $D^g(x,t)$, $D^{\bar{q}}(x,t)$ as the densities of quark,gluon,anti-quark into the initial
quark and the similar quantities for the initial anti-quark (see Fig.~\ref{Fig:D}).

Besides let us introduce $G^q(x,t)$, $G^g(x,t)$, $G^{\bar{q}}(x,t)$ as the similar densities for the initial gluon.

In complete analogy to the case of APL equations one can obtain the evolution equations for these set of densities. They are
presented in Appendix~\ref{Appendix.SetOfEquations}.

When neglecting the presence of densities of anti-quarks in gluon and quark, the combinations
$G^g(x,t)+D^g(x,t)=G(x,t)$ and $G^g(x,t)+D^g(x,t)=q(x,t)$ can be shown (see Appendix~\ref{Appendix.SetOfEquations}) to obey the equations of AP.

One of successful phenomenological model-based on the dominant role of gluon distribution $G^g$ in describing the processes with
high  hadron multiplicities \cite{Kokoulina:2004gu, Kokoulina:2005xx, Ermolov:2005tk}.

Solving the equation for $G^g$ by iteration method we see that regeneration of gluon density in the channel $G^g \to D^q \to G^g$ is associated
with small factor
\ba
    K_0 =
    \br{\frac{C_F}{2C_V}}^2
    =
    \br{\frac{4/3}{6}}^2 \approx 0.05.
    \label{K0}
\ea

Terms of such a magnitude can be neglected thus determining the accuracy of the approximation. Alternatively it
can be included as a some contribution to $K$ -factor.

Let now identify the nonsinglet structure function $q - \bar q$ with $D^q$ - a quark dominance density.
We introduce the gluon dominance density $G^g$, satisfying the equations:
\begin{figure}
    \includegraphics[width=0.85\textwidth]{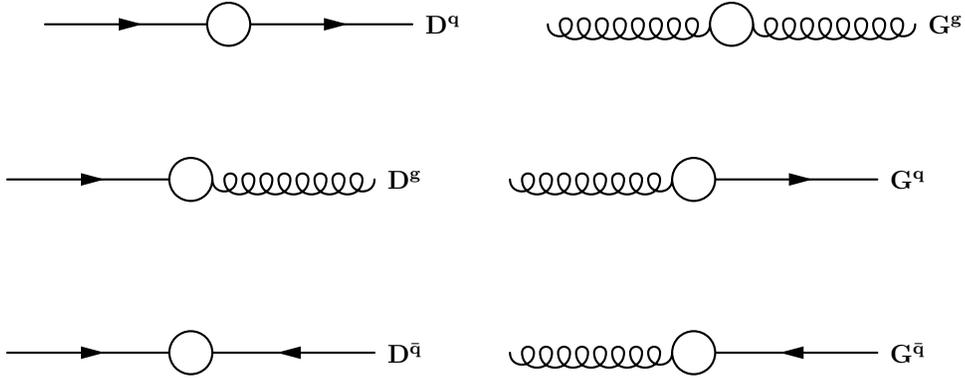}
    \caption{\label{Fig:D}
Definition of correlator densities.}
\end{figure}
\ba
    D^q\br{x,\beta_q} &=& \delta\br{x-1} +
    \int\limits_{m^2}^s \frac{\alpha_s\br{Q^2}}{2\pi} \frac{dQ^2}{Q^2}
    \int\limits_x^1 \frac{dy}{y}
    P_{qq}\br{\frac{x}{y}} D^q \br{y,\beta_{q}'},
    \label{ApproximateEquations1}
    \\
    xG^g\br{x,\beta_g} &=& \delta\br{x-1} +
    \int\limits_{m^2}^s \frac{\alpha_s\br{Q^2}}{2\pi} \frac{dQ^2}{Q^2}
    \int\limits_x^1 \frac{dy}{y}
    P_{GG}\br{\frac{x}{y}}y D_g\br{y,\beta_{g}'},
    \label{ApproximateEquations2}
\ea
where $s$ is the total invariant mass of the process (i.e. $\sqrt{s}=2E$, where $E$ is the energy of particle in beam) and
\ba
    \beta_q&=&C_F\frac{\alpha_s}{2\pi}\br{\ln\br{\frac{s}{m^2}}-1};  \qquad \,\,\,\beta_{q}'=C_F\frac{\alpha_s}{2\pi}\br{\ln\br{\frac{Q^2}{m^2}}-1};\nn \\
    \beta_g&=&2C_V\frac{\alpha_s}{2\pi}\br{\ln\br{\frac{s}{m^2}}-1}; \qquad \beta_{g}'=2C_V\frac{\alpha_s}{2\pi}\br{\ln\br{\frac{Q^2}{m^2}}-1}.
\ea
Using the solutions of the homogeneous equations for quark non-singlet density $D^q$ and $G^g$:
(we use the method similar to one developed in frames QED in \cite{Kuraev:1985hb} (see Eq. (11) and Eq. (20) in \cite{Kuraev:1985hb}):
\ba
    D^q \br{x,\beta_q} &=&
    2\beta_q \br{1-x}^{2\beta_q}\brs{\frac{1}{1-x}\br{1+\frac{3}{2}\beta_q} - \frac{1}{2}\br{1+x}} + O\br{\beta_q^2},
    \nn\\
    x G^g \br{x,\beta_g} &=&
    2\beta_g \br{1-x}^{2\beta_g}\br{\frac{1}{1-x}\br{1+\frac{11}{6}\beta_g} -2x + x^2\br{1-x}} + O\br{\beta_g^2}.
    \label{eq.SolutionsDqGg}
\ea
These solutions satisfy the properties (\ref{PProperties}), i.e.:
\ba
    \int\limits_0^1 dx~D^q\br{x,\beta_q} = 1,
    \qquad
    \int\limits_0^1 dx ~xG^g\br{x,\beta_g} = 1.
\ea
Below we consider the Drell-Yan process in collision of protons with the hard subprocess $a+b\to F^{ab}$
(where $a$ and $b$ are partons
and $F^{ab}$ is some final state produced by them) which is the part of more complicated process
$p+p \to jet_1+jet_2+F^{ab}$. Thus we choose scale of order $Q^2 = s$ (i.e. $\alpha_s\br{Q^2}=\alpha_s$)
where $\sqrt{s}=2E$ is the total energy of process in the center of mass frame.

For inclusive experiments in $pp$--collisions the cross section in center-of-mass system takes form of Drell-Yan type form:
\ba
d\sigma_{pp\to F+X}
&=&
\int\limits_0^1 dx_1
\int\limits_0^1 dy_1
\sum_{a_1}
W_{a_1} \!\!\br{x_1}
\sum_{b_1}
D_{a_1}^{b_1} \!\!\br{x_1 y_1, \beta_{b_1}} K_{a_1}
\times\nn\\
&\times&
\int\limits_0^1 dx_2
\int\limits_0^1 dy_2
\sum_{a_2}
W_{a_2} \!\!\br{x_2}
\sum_{b_2}
D_{a_2}^{b_2} \!\!\br{x_2 y_2, \beta_{b_2}} K_{a_2}
\times\nn\\
&\times& d\hat\sigma^{b_1 b_2 \to F}\!\!\br{\hat s, \hat t, \hat u}
\Theta\br{z - z_{th}},
\qquad
z=x_1 y_1 x_2 y_2,
\label{eq.csDYform}
\ea
where $W^a\br{x}$ is the probability to find a parton of sort $a$ with energy fraction $x$ inside a proton with small virtuality
(module of it's momentum square of order of $1~\GeV^2$), these quantities was obtained in \cite{Martin:2009iq} (see Appendix~\ref{Appendix.PartonDensities}).
$D_a^b\br{x,\beta}$ is the densities of parton of sort $b$ inside the parton of sort $a$.
The summation over $\brf{a_1,a_2}$ is performed over all possible partons inside the proton, i.e. $u$, $d$, $s$, $\bar u$, $\bar d$, $\bar s$, $g$.
The summation over $\brf{b_1,b_2}$ is performed over all possible partons which can be found inside the parton of sort $\brf{a_1,a_2}$, i.e.
in principle all possible partons too ($u$, $d$, $s$, $\bar u$, $\bar d$, $\bar s$, $g$).
The quantity
\ba
d\hat\sigma^{b_1 b_2 \to F}\!\!\br{\hat s, \hat t, \hat u}
\ea
in (\ref{eq.csDYform}) is the cross section of hard subprocess of two partons $b_1$ and $b_2$ fusion which actually produces the final system $F$ which is of experimental interest.
This cross section is considered already in the system of center-of-mass of this subprocess $b_1 + b_2$, i.e. these invariants $\brf{\hat s, \hat t, \hat u}$ are in this
reference frame of subprocess and have a form:
\begin{gather}
\hat s=4E^2z, \qquad \hat t=-2E^2z\br{1-\cos\hat\theta}, \qquad \hat u=-\hat s-\hat t, \nn
\end{gather}
where $z=x_1 y_1 x_2 y_2$ and $\hat\theta$ is the angle
between 3-vectors of initial parton $b_1$ momentum and the momentum of one of particles from
the created state $F$ in the center-of-mass reference frame of the subprocess, which can be expressed in terms of the
angle $\theta$ between the directions of the initial beam and the momentum of one of particles from
the created state $F$:
\begin{gather}
\cos\hat\theta = \frac{x_1 y_1 + x_2 y_2 \cos\theta}{x_2 y_2 + x_1 y_1 \cos\theta}.
\label{eq.ThetaHatThetaRelation}
\end{gather}
Here polar angle $\theta_c$ is the angle between 3-vectors of initial parton $A$ and the momentum of one of particles from
the created state $F_{AB}$ in center of mass frame of sub-process.
Polar angle $\theta$ is the angle between the directions of the initial beam and the momentum of one of particles from
the created state $F_{AB}$.

The quantities $K_{a}$ in (\ref{eq.csDYform}) are the so called $K$-factors which takes into account non-leading contribution of evolution.
The $K$-factor associated with quark density has a form \cite{Kodaira:1981nh}:
\ba
K_q=1+\frac{\alpha_s}{2\pi}k_q,
\qquad
k_q=\frac{1}{2} C_V \br{\frac{67}{18}-\frac{\pi^2}{6}} - \frac{5 n_f}{18}\approx 1.5.\nn
\ea
and $K_g$ is the $K$-factor associated with gluon density and has a form
\ba
K_g=1+\frac{1}{2}K_0
\ea
where $K_0$ is given in (\ref{K0}).

The $\Theta$-function in (\ref{eq.csDYform}) assures that experimental setup allows to register the jets of produced particles with
some finite invariant mass $s_{th}$ only, i.e. the jets with the invariant mass $s_j \geq s_{th}$. And the quantity $z_{th}$
which characterizes this threshold has the form:
\ba
    z_{th} = \frac{s_{th}}{s}.
\ea

The solutions of evolution equations for $D_a^b = D^q, G^g$ are given in (\ref{eq.SolutionsDqGg}).
The solutions for $D_a^b = D^g, D^{\bar{q}}, G^q=G^{\bar{q}}$ are presented in Appendix~\ref{Appendix.ApproximateEquations}).
Density correlators of type  $D^{q'}$ which describe the density of a quark $q'$ in quark $q$ and $D^{\bar{q}'}$ (the density
of anti-quark  $\bar{q}'$ inside the quark $q$) as well are discussed in Appendix~\ref{Appendix.ApproximateEquations}.

\section{Application to some definite subprocess}

Below we consider two types of subprocesses.
First, for the sake of demonstration we will consider the process of of associative production of top quark and the charged Higgs boson $H^-$,
since the application of our approach is more simple in this case.
And then we will use our approach to describe the experimental data from Tevatron \cite{Abe:1997yb}.

\subsection{The process $p + p \to t + H^- + jjjj$}

Let us consider now the important application of our approach to the process
\eq{
    p+p \to t + H^- + jjjj,
    \label{eq.ApplicationProcess1}
}
where $j$ denotes jet.
In this case the dominant channel of charge Higgs production is through the subprocess:
\eq{
    b + g \to t + H^-.
}
The cross section of this sub-process has the form (see (2.1) in \cite{Kidonakis:2004ib}):
\ba
    \frac{d \hat\sigma^{bg\to tH^-}}{d\cos\hat\theta} &=& \frac{\sigma_0}{\hat s}
    \left\{
        \frac{\hat s+\hat t-M_{H^-}^2}{2\hat s}- \frac{m_t^2\br{\hat u-M_{H^-}^2}+M_{H^-}^2\br{\hat t-m_t^2}+\hat s\br{\hat u-m_t^2}}{\hat s\br{\hat u-m_t^2}}\right.\nn\\
    && \qquad-
    \left.
        \frac{m_t^2\br{\hat u-M_{H^-}^2-\hat s/2}+\hat s \hat u/2}{\br{\hat u-m_t^2}^2}
    \right\},
    \label{SubprocessFortH}
\ea
where the sub-process invariants $\hat s$, $\hat t$, $\hat u$ are defined as:
\eq{
    \hat s = \br{p_b + p_g}^2,
    \qquad
    \hat t = \br{p_b + p_t}^2,
    \qquad
    \hat u = \br{p_b + p_{H^-}}^2,
    \label{eq.SubProcessInvariants1}
}
and angle $\hat \theta$ is the angle between the momenta of initial $b$--quark and produced $t$--quark in the
reference frame of center of mass of subprocess
(i.e. $\vv{p_b} = -\vv{p_g}$ where $\vv{p_b}$ and $\vv{p_g}$ are the 3-momenta of initial $b$-quark and the gluon correspondingly);
the quantity $\sigma_0$ is the following constant
\ba
    \sigma_0 = \frac{\pi \alpha \alpha_s\br{m_b^2 \tan^2\beta + m_t^2 \cot^2\beta}}{6 \, M_W^2 \sin^2\theta_W},
\ea
where $\theta_W$ is the Weinberg angle and $\beta$ is the parameter of
Minimal Supersymmetric Standard Model (MSSM). For $\tan\beta=40$ and $\alpha_s=0.1$ we have $\sigma_0 \approx 0.06$.

\begin{figure}
    \includegraphics[width=0.9\textwidth]{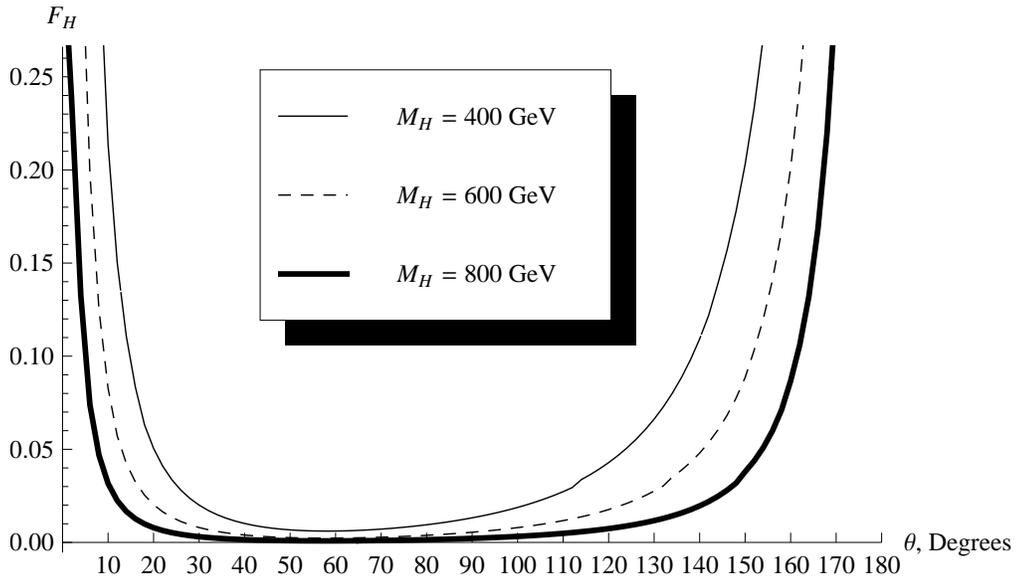}
    \caption{\label{Fig:tHtheta}
            The angular dependence of quantity $F_H$ defined in (\ref{eq.FH}).
        }
\end{figure}

\begin{figure}
    \includegraphics[width=0.9\textwidth]{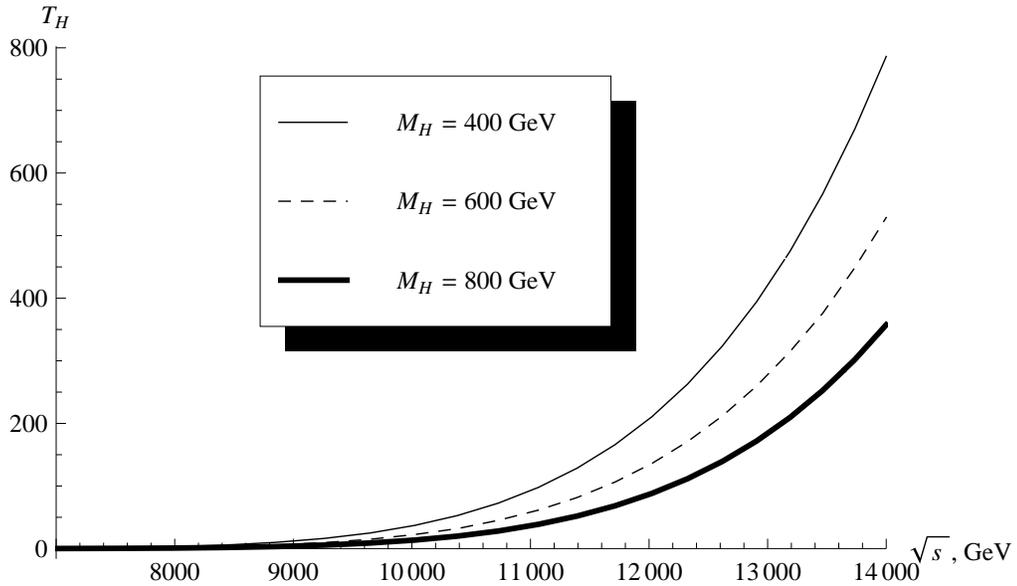}
    \caption{\label{Fig:tHs}
        The quantity $T_H$ defined in (\ref{eq.TH}) as function of total invariant mass $s$.
    }
\end{figure}

The application of our master--formula (\ref{eq.csDYform}) to the process (\ref{eq.ApplicationProcess1}) gives the cross section in
the following form:
\eq{
&\frac{d\sigma_{pp\to t H^-+jjjj}}{d\cos\theta}
=
\int\limits_0^1 dx_1
\int\limits_0^1 dy_1
\int\limits_0^1 dx_2
\int\limits_0^1 dy_2
\,\,\theta\br{z - z_{th}}
\frac{d \hat\sigma^{bg\to tH^-}}{d\cos\theta}
\times\nn\\
&\quad\times
    \br{
        W_u \br{x_1} D^{q'} \!\!\br{x_1 y_1, \beta_q} K_q
        +
        W_d \br{x_1} D^{q'} \!\!\br{x_1 y_1, \beta_q} K_q
        +
        W_g \br{x_1} G^q \!\br{x_1 y_1, \beta_g} K_g
    }
\times
\nn\\
&\quad\times
    \br{
        \frac{}{}
        W_u \br{x_2} D^g \!\br{x_2 y_2, \beta_q} K_q
        +
        W_d \br{x_2} D^g \!\br{x_2 y_2, \beta_q} K_q
        +
        W_g \br{x_2} G^g \!\br{x_2 y_2, \beta_g} K_g
    }.
\label{eq.csDYtH0}
}
At this stage we need to notice that the cross section of sub-process in (\ref{eq.csDYtH0})
depends on angle $\theta$ between the direction of momentum of produced $t$-quark and the beam in the system of center-of-mass of initial
proton-proton beams, while the expression (\ref{SubprocessFortH})
depends on the scattering angle $\hat \theta$ in the center-of-mass reference frame of sub-process.
Since these angles correspond to each other with the relation (\ref{eq.ThetaHatThetaRelation}),
i.e.:
\eq{
    d\cos\hat\theta =
    \frac{x_2^2 y_2^2-x_1^2 y_1^2}{\br{x_2 y_2 + x_1 y_1 \cos\theta}^2}
    \,
    d\cos\theta,
}
we obtain the following final form of the cross section (\ref{eq.csDYtH0}):
\eq{
&\frac{d\sigma_{pp\to t H^-+jjjj}}{d\cos\theta}
=
\int\limits_0^1 dx_1
\int\limits_0^1 dy_1
\int\limits_0^1 dx_2
\int\limits_0^1 dy_2
\,\,\theta\br{z - z_{th}}
\frac{d \hat\sigma^{bg\to tH^-}}{d\cos\hat\theta}
\frac{x_2^2 y_2^2-x_1^2 y_1^2}{\br{x_2 y_2 + x_1 y_1 \cos\theta}^2}
\times\nn\\
&\quad\times
    \br{
        W_u \br{x_1} D^{q'} \!\!\br{x_1 y_1, \beta_q} K_q
        +
        W_d \br{x_1} D^{q'} \!\!\br{x_1 y_1, \beta_q} K_q
        +
        W_g \br{x_1} G^q \!\br{x_1 y_1, \beta_g} K_g
    }
\times
\nn\\
&\quad\times
    \br{
        \frac{}{}
        W_u \br{x_2} D^g \!\br{x_2 y_2, \beta_q} K_q
        +
        W_d \br{x_2} D^g \!\br{x_2 y_2, \beta_q} K_q
        +
        W_g \br{x_2} G^g \!\br{x_2 y_2, \beta_g} K_g
    }.
\label{eq.csDYtH}
}
The dependence of this cross section on scattering angle $\theta$
is presented in Fig.~\ref{Fig:tHtheta} via quantity
\ba
F_H \br{\theta} = \frac{1}{\sigma_0} \frac{d\sigma_{pp\to t H^- + jjjj}}{d\cos\theta},
\label{eq.FH}
\ea
which is built for few values of charges Higgs boson mass $M_H$.
The total cross section of this process is proportional
to the quantity $T_H$:
\ba
    T_H \br{s} = \int\limits_0^\pi d\theta \, F_H \br{\theta},
    \label{eq.TH}
\ea
which is presented in Fig.~\ref{Fig:tHs}.
%

\subsection{Six-jets production at Tevatron}

In paper \cite{Abe:1997yb} the data for six-jet production are present.
The application of master--formula (\ref{eq.csDYform}) to this process gives
more complicated result since we need take into account few subprocesses
\cite{Nakamura:2010zzi}:
\ba
\frac{d\sigma(q\bar{q}\to q\bar{q})}{d\cos\hat\theta}&=&\frac{\alpha_s^2}{9\hat s}
\brs{\frac{\hat t^2+\hat u^2}{\hat s^2}+\frac{\hat s^2+\hat u^2}{\hat t^2}-\frac{2\hat u^2}{3\hat s\hat t}}, \nn \\
\frac{d\sigma(q\bar{q}'\to q\bar{q}')}{d\cos\hat\theta} &=& \frac{\alpha_s^2}{9\hat s}\frac{\hat t^2+\hat u^2}{\hat s^2},  \nn \\
\frac{d\sigma(g g \to q\bar{q})}{d\cos\hat\theta} &=& \frac{\alpha_s^2}{24\hat s}\br{\hat t^2+\hat u^2}
\br{\frac{1}{\hat t \hat u}-\frac{9}{4\hat s^2}}, \nn \\
\label{subprocess}
\ea
where $\hat\theta$ is the angle between the direction of motion of initial parton and the momentum of final quark
and invariant $\hat s$, $\hat t$, $\hat u$ are defined in the same manner as in (\ref{eq.SubProcessInvariants1}).

\begin{figure}
    \includegraphics[width=0.9\textwidth]{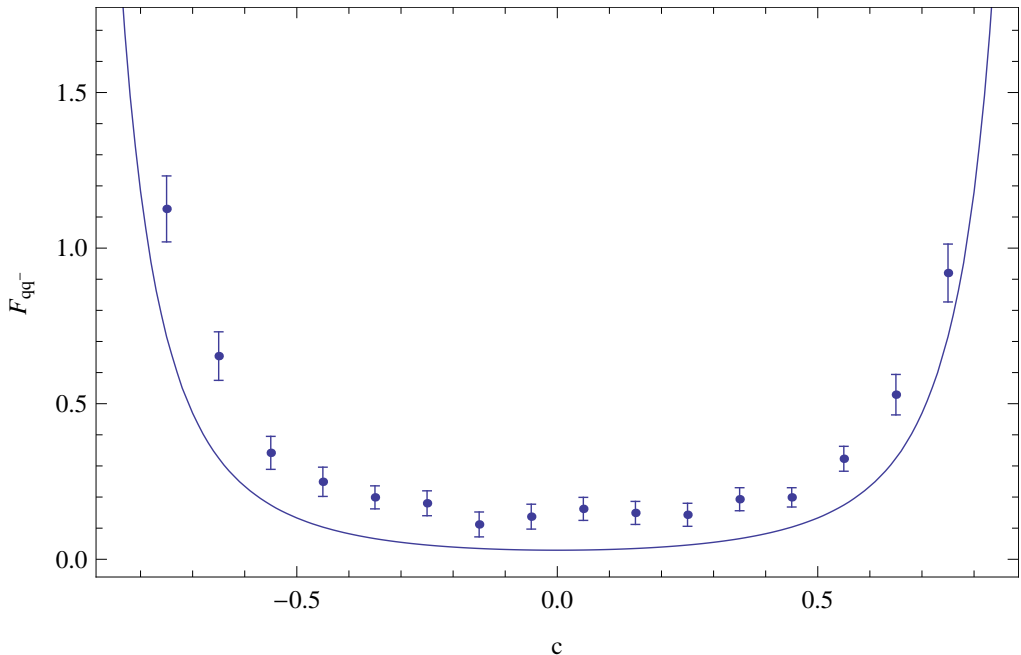}
    \caption{\label{Fig:qqtheta}
            The angular dependence of quantity $F_{qq}$ defined in (\ref{eq.Fqq}) for different initial total energy $\sqrt{s}$.
        }
\end{figure}

The comparison of angular distribution of jets momenta in proton--proton scattering from (\ref{subprocess})
\ba
    F_{qq} \br{\theta} = \frac{1}{\sigma_\text{tot}} \frac{d\sigma_{pp\to 6 j}}{d\cos\theta},
    \label{eq.Fqq}
\ea
(where $\sigma_\text{tot}$ is the total cross section of 6 jet production)
with the experimental results (see figure 6 (a) in paper \cite{Abe:1997yb})
is shown in Fig.~\ref{Fig:qqtheta} for different initial total energy $\sqrt{s}$.

\section{Conclusion}

In this paper we discuss some modification of the method of taking into account the QCD leading logarithm
radiative corrections based on the Structure Functions approach.
Modification consists in construction of set of evolution equations for density of parton of sort $a$ in the initial quark
$D^a\br{x,\beta_q}$ and density of parton of sort $b$ to be in the initial gluon $G^b\br{x,\beta_g}$.
This set
of equations is solved in quark and gluon dominance approximation $D^q \gg D^a$, $a \neq q$ and $G^g \gg G^a$, $a \neq g$.
This approximation can be improved for the accuracy level which is required by using the iteration procedure.
This assumption is known as a gluon dominance which is used in
description of multiplicity of $\pi$--mesons in the hadron collisions
\cite{Kokoulina:2004gu, Kokoulina:2005xx, Ermolov:2005tk}.

We present approximate solution for $D^a$, $G^a$
and demonstrate the application of this function to the problem of calculation of QCD radiative
correction calculation in some particular processes.

Some efforts was paid to the problem of calculation of subprocesses cross section in the next--to--leading
approximation. Main attention was paid to the 2-loops level contributions in \cite{Kidonakis:2000ui, Kidonakis:2003qe}.
As a result the terms of order $(\alpha_s L^2)$, $(\alpha_s L^2)^2$ was taken into account.
However the emission of real (soft and hard) gluons with the 1-loop radiative corrections was not considered.
The role of radiative (virtual and real) corrections leads to the change of the $\br{\alpha L^2}^n$--regime
to a single-logarithmical regime (i.e. only terms $\sim \br{\alpha L}^n$ remains) in inclusive experimental approach.
Single-logarithmical approach is determined by renormalization group evolution equation and thus allows us to use
the Structure Function approach to obtain the cross section in leading (i.e. $\br{\alpha L}^n$) and
next--to--leading (i.e. $\alpha \br{\alpha L}^n$) approximation.

Functions $W^a(x)$ describe the probability to find parton $a$ inside a proton with off mass shell about one $\GeV $ squared.
These functions was builded in \cite{Martin:2009iq} as a result of self-consistent analysis of many sub-processes,
and besides shown to satisfy the momentum and number sum rules.
For experimental setups with the product of subprocess detected which moves at large angles
with invariant mass square exceeding some threshold value $s_{th}=s \,z_{th}$ the role of
"sea"--partons in the proton can be neglected.

\acknowledgments
One of us (EAK) is grateful to DESY theoretical group and to Dr. A. Ali for valuable discussions when this paper was started.
This work was supported by the Heisenberg--Landau program, grant HLP-2012-11.

\appendix

\section{Full set of evolution equations}
\label{Appendix.SetOfEquations}

Let introduce three distributions $D^a=D_q^a(y,t)$ with $a=q,g,\bar{q}$ which describe the number of partons $a$ with energy fraction $y$
inside the parent quark. Similarly one must introduce three quantities $\bar{D}^a$ and three distributions $G^a$.
Keeping in mind the absence of transition of a quark (antiquark) to antiquark (quark) $P_{q\bar{q}}=P_{\bar{q}q}=0$ in lowest order of
perturbation theory, the evolution equations of these 3 sets of distributions will have a form similar to ones for quark and gluon densities
inside a proton given above. Similar consideration was used in frames of QED in paper \cite{Arbuzov:2010zzb}.
For quark densities
\ba
\frac{d}{d t}D^q(x,t)&=&\frac{\alpha(t)}{2\pi}\int\limits_x^1\frac{d y}{y}\brs{D^q(y,t)P_{qq}\br{\frac{x}{y}}+D^g(y,t)P_{qg}\br{\frac{x}{y}}}; \nn \\
\frac{d}{d t}D^{\bar{q}}(x,t)&=&\frac{\alpha(t)}{2\pi}\int\limits_x^1\frac{d y}{y}\brs{D^{\bar{q}}(y,t)P_{qq}\br{\frac{x}{y}}+D^g(y,t)P_{qg}\br{\frac{x}{y}}}; \nn \\
\frac{d}{d t}D^g(x,t)&=&\frac{\alpha(t)}{2\pi}\int\limits_x^1\frac{d y}{y}\brs{D^g(y,t)P_{gg}\br{\frac{x}{y}}+D^{\bar{q}}(y,t) P_{gq}\br{\frac{x}{y}}+D^q(y,t)P_{gq}\br{\frac{x}{y}}}.
\ea
Similar set with the replacement $D^a\to \bar{D}^a$ take place for anti-quark densities. For gluon densities we have
\ba
\frac{d}{d t}G^{\bar{q}}(x,t)&=&\frac{d}{d t}G^q(x,t)=\frac{\alpha(t)}{2\pi}\int\limits_x^1\frac{d y}{y}\brs{G^q(y,t)P_{qq}\br{\frac{x}{y}}+G^g(y,t)P_{qg}\br{\frac{x}{y}}}; \nn \\
\frac{d}{d t}G^g(x,t)&=&\frac{\alpha(t)}{2\pi}\int\limits_x^1\frac{d y}{y}\brs{G^g(y,t)P_{gg}\br{\frac{x}{y}}+G^{\bar{q}}(y,t) P_{gq}\br{\frac{x}{y}}+G^q(y,t)P_{gq}\br{\frac{x}{y}}}.
\ea

It follows from these sets of equations
\ba
\frac{d}{d t}\br{D^g(x,t)+G^g(x,t)}&=&\frac{\alpha(t)}{2\pi}\int\limits_x^1\frac{d y}{y}
\left[\br{D^g(y,t)+G^g(y,t)}P_{gg}\br{\frac{x}{y}}+ \right.\nn \\
&+&\left.
\br{D^{\bar{q}}(y,t)+G^{\bar{q}}(y,t)} P_{gq}\br{\frac{x}{y}}+\br{D^q(y,t)+G^q(y,t)}P_{gq}\br{\frac{x}{y}}\right]; \nn\\
\frac{d}{d t}\br{D^q(x,t)+G^q(x,t)}&=&\frac{\alpha(t)}{2\pi}\int\limits_x^1\frac{d y}{y}
\left[\br{D^q(y,t)+G^q(y,t)}P_{qq}\br{\frac{x}{y}}+ \right. \nn \\
&+&\left.
\br{D^g(y,t)+G^g(y,t)} P_{qg}\br{\frac{x}{y}}\right].\nn
\ea
When omitting the densities of anti-quarks $D^{\bar{q}}$ and $G^{\bar{q}}$ inside the quark and the gluon and identifying
\ba
D^q+G^q&=&q, \nn\\
D^g+G^g&=&G, \nn
\ea
we reproduce Altarelli--Parisi equations (\ref{DGLAP}).

Our statement about numerical smallness of the contribution of the intermediate quark (anti-quark) states in the evolution of gluon density follows from iteration
procedure in solving the first equation of gluon set. So it can be taken into account by including as a relevant contribution to $K$-factor.
Besides only light quarks must be considered, describing the experiments without quark jets production.

\section{Approximate evolution equations}
\label{Appendix.ApproximateEquations}

A quark dominance consist in suggestion $D^q \gg \bar{D}^q,G^q$. Gluon dominance imply $G^g \gg D^g,\bar{D}^g$ and besides $D^g \gg D^{\bar{q}}$.
Set of equations in these approximations reads as
\ba
\frac{d}{d t}D^q(x,t)&=&\frac{\alpha(t)}{2\pi}\int\limits_x^1\frac{d y}{y}D^q(y,t)P_{qq}\br{\frac{x}{y}}; \nn \\
\frac{d}{d t}D^{\bar{q}}(x,t)&=&\frac{\alpha(t)}{2\pi}\int\limits_x^1\frac{d y}{y}D^g(y,t)P_{qg}\br{\frac{x}{y}}; \nn \\
\frac{d}{d t}D^g(x,t)&=&\frac{\alpha(t)}{2\pi}\int\limits_x^1\frac{d y}{y}D^q(y,t)P_{gq}\br{\frac{x}{y}}; \nn \\
\frac{d}{d t}G^q(x,t)=\frac{d}{d t}G^{\bar{q}}(x,t)&=&\frac{\alpha(t)}{2\pi}\int\limits_x^1\frac{d y}{y}G^g(y,t)P_{qg}\br{\frac{x}{y}}; \nn \\
\frac{d}{d t}G^g(x,t)&=&\frac{\alpha(t)}{2\pi}\int\limits_x^1\frac{d y}{y}G^g(y,t)P_{gg}\br{\frac{x}{y}}.
\ea
Note that the equation for $D^q$ coincide  with equation for non-singlet quark density $q_{NS}=q-\bar{q}$.
The equations for $D^q$ and $G^g$ are given above.

Keeping in mind the solution of evolution equations
\ba
\frac{\partial}{\partial t} A(x,\beta_{qt})&=&C_F\frac{\alpha_{qt}}{2\pi}\brs{a\beta_{qt}+b\beta_{qt}^2+...}=\frac{d\beta_{qt}}{d t}\brs{a\beta_{qt}+b\beta_{qt}^2+...}; \nn \\
A(x,\beta_q)&=&a\frac{1}{2}\beta_q^2+b\frac{1}{3}\beta_q^3+...,
\ea
we obtain
\ba
D^g(x,\beta_q)=G^q(x,\beta_q)&=&\frac{1}{2}\br{1+(1-x)^2}\beta_q^2+O(\beta_q^3); \nn \\
D^{\bar{q}}(x,\beta_q)=D^{q'}(x,\beta_q)=D^{\bar{q}'}(x,\beta_q)&=&\frac{1}{2}\phi(x)\beta_q^2+O(\beta_q^3), \nn \\
\phi(x)&=&\frac{1}{3 x}(1-x)(4+7x+4x^2)+2(1+x)\ln x.
\ea

\section{Parton densities in the proton}
\label{Appendix.PartonDensities}

Keeping in mind the problem of description of inelastic processes in high energy collision of protons
it seems to be naturally consider proton as an objects with definite contents from quarks and gluons.
It implied the presence of the preliminary evolution from mass shell to virtuality of order $1~\GeV^2$
of all the constituents of proton.

Note that due to condition $x_1x_2>z_{th}$ only valence quark and gluons inside proton take part in
Drell-Yan process.
We will choice the density of the valence quarks and gluons
approximately as ones found in paper \cite{Martin:2009iq}:
\ba
xW_u(x)&=&A_ux^{\eta_1}(1-x)^{\eta_2}(1+\epsilon_u \sqrt{x}+\gamma_u x); \nn \\
A_u&=&0.2; \,\,\,\eta_1=-0.73; \,\,\,\eta_2=3.3; \nn \\
xW_d(x)&=&A_dx^{\eta_3}(1-x)^{\eta_4}(1+\epsilon_d \sqrt{x}+\gamma_d x); \nn \\
A_d&=&18; \,\,\,\eta_3=0,1; \,\,\,\eta_4=6; \nn \\
xW_g(x)&=&A_gx^{\delta_g}(1-x)^{\eta_g}(1+\epsilon_g \sqrt{x}+\gamma_g x)+
A_{g'}x^{\delta_{g'}}(1-x){\eta_{g'}}, \nn \\
A_g&=&0.0012216.
\ea
Numerical constants must be chosen in such a way to satisfy the constrains from number sum rules
\ba
\int\limits_0^1 d x W_u(x)=2; \qquad
\int\limits_0^1 d x W_d(x)=1,
\ea
and besides the momentum sum rule
\ba
\int\limits_0^1 d x x \brs{W_u(x) + W_d(x) + W_g(x)+S(x)}=1,
\ea
with $S(x)$ is the sea contribution. \\
Where
$\gamma_u = 8.9924$, $\gamma_d = 7.4730$, $\eta_g = 2.3882$, $\eta_{g'} =0$,
$A_{g'}=0$, $\delta_g=-0.83657$, $\delta_{g'}=0$, $\gamma_g =1445.5$, $\epsilon_u = -2.3737$,
$\epsilon_d = -4.3654$, $\epsilon_g = -38.997$.
More complicated expressions for densities which was extracted from description of
fixed target HERA and Tevatron experiments are presented in \cite{Martin:2009iq}.
%

\bibliographystyle{D:/Physics/Bibliography/Styles/h-physrev5}
\bibliography{D:/Physics/Bibliography/Main}

\end{document}